\pdfoutput=1

\documentclass[11pt,twoside,a4paper,cmspaper,final,collab]{cms-tdr}
\def\svnVersion{81602:81824}\def\cmsCernNo{2011-169}\def\cmsCernDate{\today}\def\cmsMessage{Submitted to Physical Review D}

\begin{document}\cmsNoteHeader{EWK-10-010}

\hyphenation{had-ron-i-za-tion}
\hyphenation{cal-or-i-me-ter}
\hyphenation{de-vices}
\RCS$Revision: 81824 $
\RCS$HeadURL: svn+ssh://alverson@svn.cern.ch/reps/tdr2/papers/EWK-10-010/trunk/EWK-10-010.tex $
\RCS$Id: EWK-10-010.tex 81824 2011-10-22 14:43:07Z alverson $
\newcommand{\qt}{\ensuremath{q_{\mathrm{T}}}\xspace}
\providecommand{\POWHEG}{\textsc{powheg}\xspace}
\providecommand{\FEWZ}{\textsc{fewz}\xspace}
\providecommand{\POWHEGPYTHIA}{\textsc{powheg+pythia}\xspace}
\newlength{\cmsfigwid}
\ifthenelse{\boolean{cms@external}}{\setlength\cmsfigwid{0.8\columnwidth}}{\setlength\cmsfigwid{0.45\textwidth}}
\newlength{\cmsLargefigwid}
\ifthenelse{\boolean{cms@external}}{\setlength\cmsLargefigwid{0.9\columnwidth}}{\setlength\cmsLargefigwid{0.9\textwidth}}
\ifthenelse{\boolean{cms@external}}{\newcommand{\xleft}{top\xspace}}{\newcommand{\xleft}{left\xspace}}
\ifthenelse{\boolean{cms@external}}{\newcommand{\xright}{bottom\xspace}}{\newcommand{\xright}{right\xspace}}
\cmsNoteHeader{EWK-10-010} 
\title{\texorpdfstring{Measurement of the Rapidity and Transverse Momentum Distributions of $\cPZ$ Bosons in \Pp\Pp\ Collisions at $\sqrt{s}=7$ TeV}{Measurement of the Rapidity and Transverse Momentum Distributions of Z Bosons in pp Collisions at sqrt(s)=7 TeV}}

\date{\today}

\abstract{ Measurements of the normalized rapidity ($y$) and
  transverse momentum ($q_\mathrm{T}$) distributions of Drell--Yan
  muon and electron pairs in the \cPZ-boson mass region ($60<M_{\ell\ell}<120\,\mathrm{GeV}$) are reported.  The results are
  obtained using a data sample of proton-proton collisions at a
  center-of-mass energy of 7~TeV, collected by the CMS experiment at
  the LHC, corresponding to an integrated luminosity of 36 pb$^{-1}$.
  The distributions are measured over the ranges $|y|<3.5$ and
  $q_\mathrm{T}<600\,\mathrm{GeV}$ and compared with QCD calculations
  using recent parton distribution functions.  Overall agreement is
  observed between the models and data for the rapidity distribution,
  while no single model describes the \cPZ\ transverse-momentum
  distribution over the full range.  }

\hypersetup{%
pdfauthor={CMS Collaboration},%
pdftitle={Measurement of the Rapidity and Transverse Momentum Distributions of Z Bosons in pp Collisions at sqrt(s)=7 TeV},%
pdfsubject={CMS},%
pdfkeywords={CMS, Z, PDF, transverse-momentum, rapidity}}

\maketitle 

\section{Introduction}

The production of $\cPZ$ and $\PW$ bosons, which may be identified through
their leptonic decays, is theoretically well described within the
framework of the standard model.  Total and differential cross
sections have been calculated to next-to-next-to-leading-order (NNLO)
\cite{Anastasiou:2003ds,Melnikov:2006kv}.  The dominant
uncertainties in the calculation arise from imperfect knowledge of the
parton distribution functions (PDFs), from the uncertainty in the
strong-interaction coupling $\alpha_{s}$, and from the choice of
quantum chromodynamics (QCD) renormalization and factorization scales.
Measurements of the inclusive $\cPZ$ and $\PW$ production cross sections
performed by the Compact Muon Solenoid (CMS)
experiment~\cite{CMS-VBTF-2} show agreement with the latest
theoretical predictions both for the absolute value and for the ratios
$\PWp/\PWm$ and \PW/\cPZ.  Likewise, agreement is found for the
measurement of the dilepton mass distribution over a wide range
\cite{CMS-DSDM}.

In this paper, we present measurements of the normalized rapidity and
transverse momentum distributions for Drell--Yan muon and electron
pairs in the \cPZ-boson mass region ($60 < M_{\ell\ell} <
120\GeV$).  The results are obtained from a sample of
proton-proton collisions at a center-of-mass energy of 7~\TeV,
recorded by the CMS detector at the Large Hadron Collider (LHC) in
2010, and corresponding to an integrated luminosity of $35.9 \pm
1.4\pbinv$.  The measurement of the rapidity ($y$) and
transverse momentum (\qt) distributions of the $\cPZ$ boson provides new
information about the dynamics of proton collisions at high energies.
The $y$ distribution of $\cPZ$ bosons is sensitive to the PDFs,
particularly when measured in the forward region ($|y|>2.5$), as done
in this paper.  The \qt spectrum provides a better understanding of
the underlying collision process at low transverse momentum, and tests
NNLO perturbative QCD predictions at high transverse momentum.

The rapidity is defined as \(y=\frac{1}{2} \ln \left[\left(
    E+q_L\right)/\left(E-q_L\right)\right] \), where $E$ is the energy of the \cPZ-boson candidate
and $q_L$ is its longitudinal momentum along
the anti-clockwise beam axis (the $z$ axis of the detector).  The
Z-boson $y$ and $\qt$ are determined from the lepton momenta, which
can be measured with high precision in the CMS detector.  The measured
differential dimuon and dielectron cross sections are normalized to
the inclusive $\cPZ$ cross section, thereby canceling several
sources of systematic uncertainties.

The \cPZ-boson $y$ and \qt distributions have been measured by the
Tevatron experiments \cite{Abazov:dsdptee_2010D0,
  Abazov:dsdptmm_2010D0, Aaltonen:dsdy_2010CDF,
  Abazov:dsdy_2007D0}. In this paper, we report measurements which
cover the range in rapidity up to 3.5 and in transverse momentum up to
600\GeV, a similar range to results recently reported by the ATLAS
experiment~\cite{DSDPT-ATLAS, NEW-ATLAS}.  The rapidity measurement is
sensitive to the PDFs for proton momentum fractions ($x$) between
$4\times 10^{-4}$ and $0.43$.

\section{The CMS detector}
\label{sec:CMS}
\par
The central feature of the CMS apparatus is a superconducting solenoid
of 6~m internal diameter, providing a magnetic field of $3.8$~T.
Within the field volume are a silicon pixel and strip tracker, a
crystal electromagnetic calorimeter (ECAL), and a brass/scintillator
hadron calorimeter (HCAL).  The inner tracker measures charged
particle trajectories in the pseudorapidity range $|\eta| < 2.5$ and
provides a transverse momentum ($\pt$) resolution of about 1--2\% for
charged particles with $\pt$ up to 100~\GeV.  The pseudorapidity
$\eta$ is defined as $\eta=-\ln(\tan(\theta/2))$, where $\theta$ is
the polar angle with respect to the anti-clockwise beam direction.
The electromagnetic calorimeter contains nearly $76\,000$
lead-tungstate crystals that provide a coverage of $|\eta| < 1.48$ in
a cylindrical barrel region and of $1.48 < |\eta| < 3.0$ in two endcap
regions.  The ECAL has an energy resolution of better than $0.5\%$ for
unconverted photons with transverse energies above $100~\GeV$.  The
energy resolution is $3\%$ or better for electrons with $|\eta|<2.5$.
The regions ($3.0 < |\eta| < 5.0$) are covered by sampling Cherenkov
calorimeters (HF) constructed with iron as the passive material and
quartz fibers as the active material.  The HF calorimeters have an
energy resolution of about $10\%$ for electron showers.  Muons are
detected in the range $|\eta|< 2.4$, with detection planes based on
three technologies: drift tubes, cathode strip chambers, and resistive
plate chambers.  Matching segments from the muon system to tracks
measured in the inner tracker results in a \pt resolution of between 1
and 5\% for muons with \pt up to 1\TeV.  Data are selected online
using a two-level trigger system. The first level, consisting of
custom hardware processors, selects events in less than 1~$\mu$s,
while the high-level trigger processor farm further decreases the
event rate from around 100~kHz to about 300~Hz before data storage.  A
more detailed description of CMS can be found in Ref.~\cite{JINST}.
\par

\section{Analysis procedure, data samples, and event selection}\label{sec:procedure}

The differential cross section is determined in each $y$ or \qt bin by
subtracting from the number of detected events in a bin the estimated
number of background events.  The distributions are corrected for
signal acceptance and efficiency and for the effects of detector
resolution and electromagnetic final-state radiation
(FSR) using an unfolding technique based on the inversion of a response
matrix.  The final result takes into account the bin width and is
normalized by the measured total cross section.

The measurements of the rapidity and transverse momentum spectra are
based on samples of over $12\,000$ \cPZ-boson events reconstructed in
each dilepton decay mode, and collected using high \pt single-lepton
triggers.  The lepton identification requirements used in the analysis
are the same as those employed in the measurement of the inclusive $\PW$ and $\cPZ$ cross sections~\cite{CMS-VBTF-1}.  For the \cPZ-boson candidates
selected, the pairs of leptons, $\ell$, are required to have a
reconstructed invariant mass in the range $60 < M_{\ell\ell} <
120\GeV$.

Muon events are collected using a trigger requiring a single muon,
with a \pt threshold that was increased from 9 to $15\,\GeV$ in
response to increasing LHC luminosity during the data-taking
period. The two muon candidates with the highest \pt in the event are
used to reconstruct a \cPZ-boson candidate. Muons are required to have
$\pt>20\GeV$, $|\eta|<2.1$, and to satisfy the standard CMS muon
identification criteria described in Ref.~\cite{CMS-VBTF-1}.  In
addition, the two muons are required to be isolated by calculating the
sum of additional track momenta ($I_\mathrm{trk}$) and hadron
calorimeter energy not associated with the muon ($I_\mathrm{HCAL}$) in
a cone $\Delta R = \sqrt{(\Delta\eta)^2+(\Delta\phi)^2} < 0.3$ around
the muon momentum direction, and requiring
$(I_\mathrm{trk}+I_\mathrm{HCAL})/\pt(\mu)< 0.15$.  Information from
the ECAL is not used as a criterion for isolation to avoid
dependencies on FSR modeling \cite{CMS-DSDM}.  At least one of the
reconstructed muons must have triggered the event.  The two muons in a
pair are required to have opposite charges as determined by track
curvature.  The invariant mass distribution for selected events is
shown in Fig.~\ref{fig:massplots}.  We compare the kinematic
distributions from the data and the simulations described below, and
find that they agree.

\ifthenelse{\boolean{cms@external}}{\newcommand{\figCaption}{Dilepton invariant mass distributions for the muon channel
  (top) and the electron channel (middle and bottom).
  The the upper electron plot shows the invariant mass
  distribution for events with both electrons in the ECAL and the lower
  for events with one electron in the ECAL and the
  other in the HF.  Each plot shows the data observation
  compared to the signal as predicted by \POWHEG on top of the background estimated
  from a combination of simulation and data.  The background is very low in the
  muon channel.}}{\newcommand{\figCaption}{Dilepton invariant mass distributions for the muon channel
  (top) and the electron channel (bottom).
  The bottom-left plot for the electron shows the invariant mass
  distribution for events with both electrons in the ECAL and the bottom-right
  for events with one electron in the ECAL and the
  other in the HF.  Each plot shows the data observation
  compared to the signal as predicted by \POWHEG on top of the background estimated
  from a combination of simulation and data.  The background is very low in the
  muon channel.}}
\begin{figure}[tbhp]
\begin{center}
\includegraphics[width=\cmsfigwid]{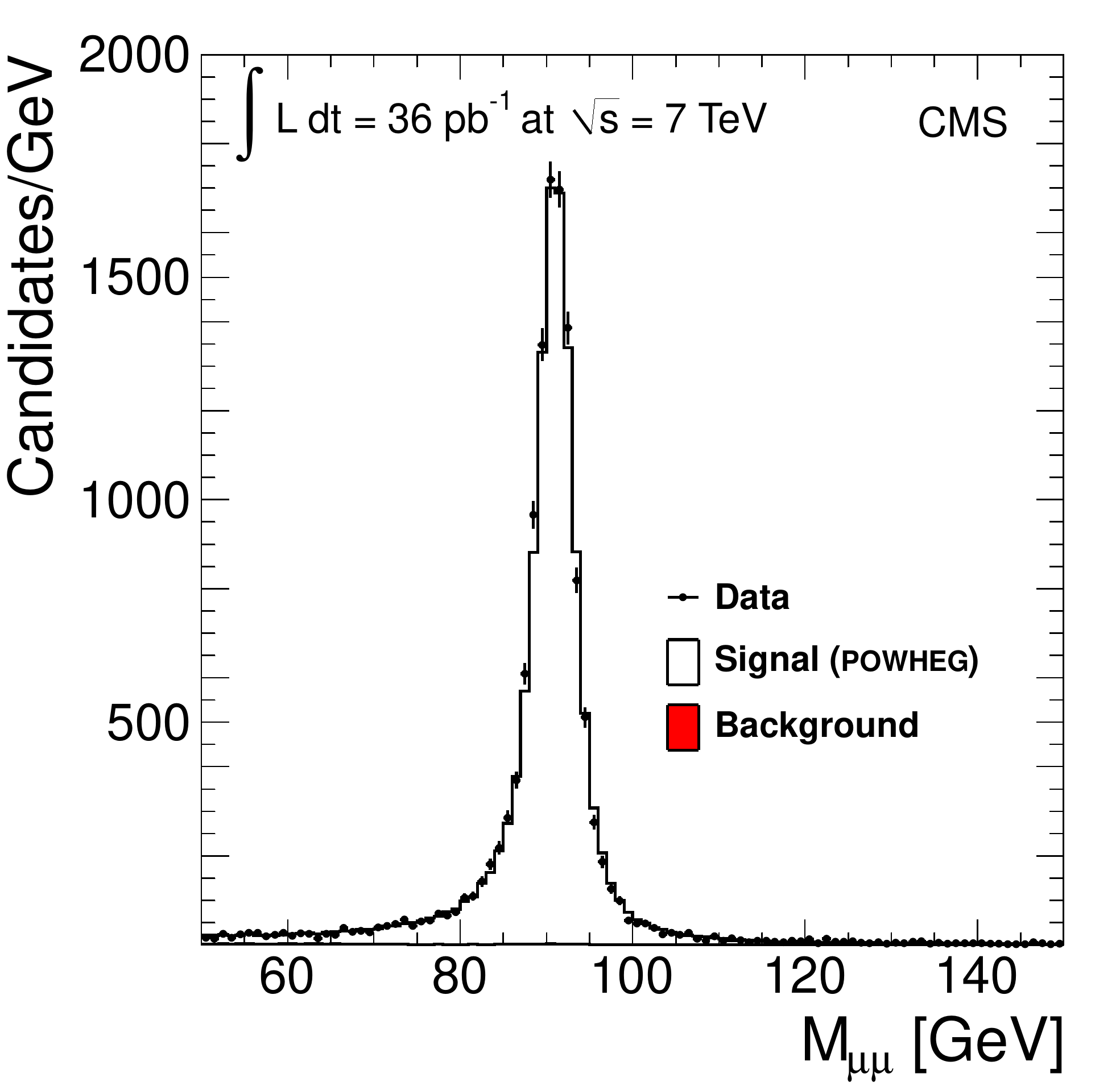}
\\
\includegraphics[width=\cmsfigwid]{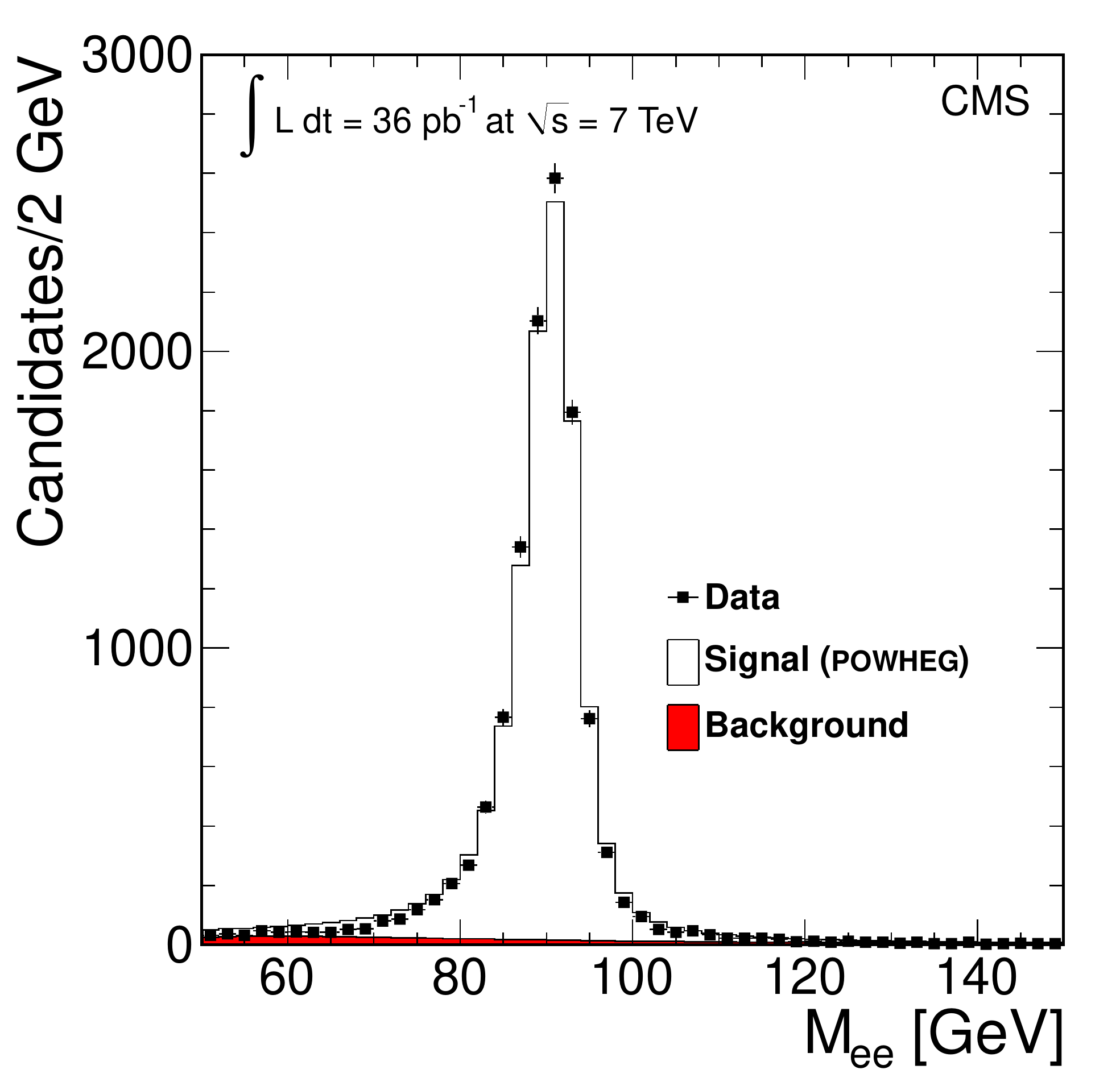}
\hfill
\includegraphics[width=\cmsfigwid]{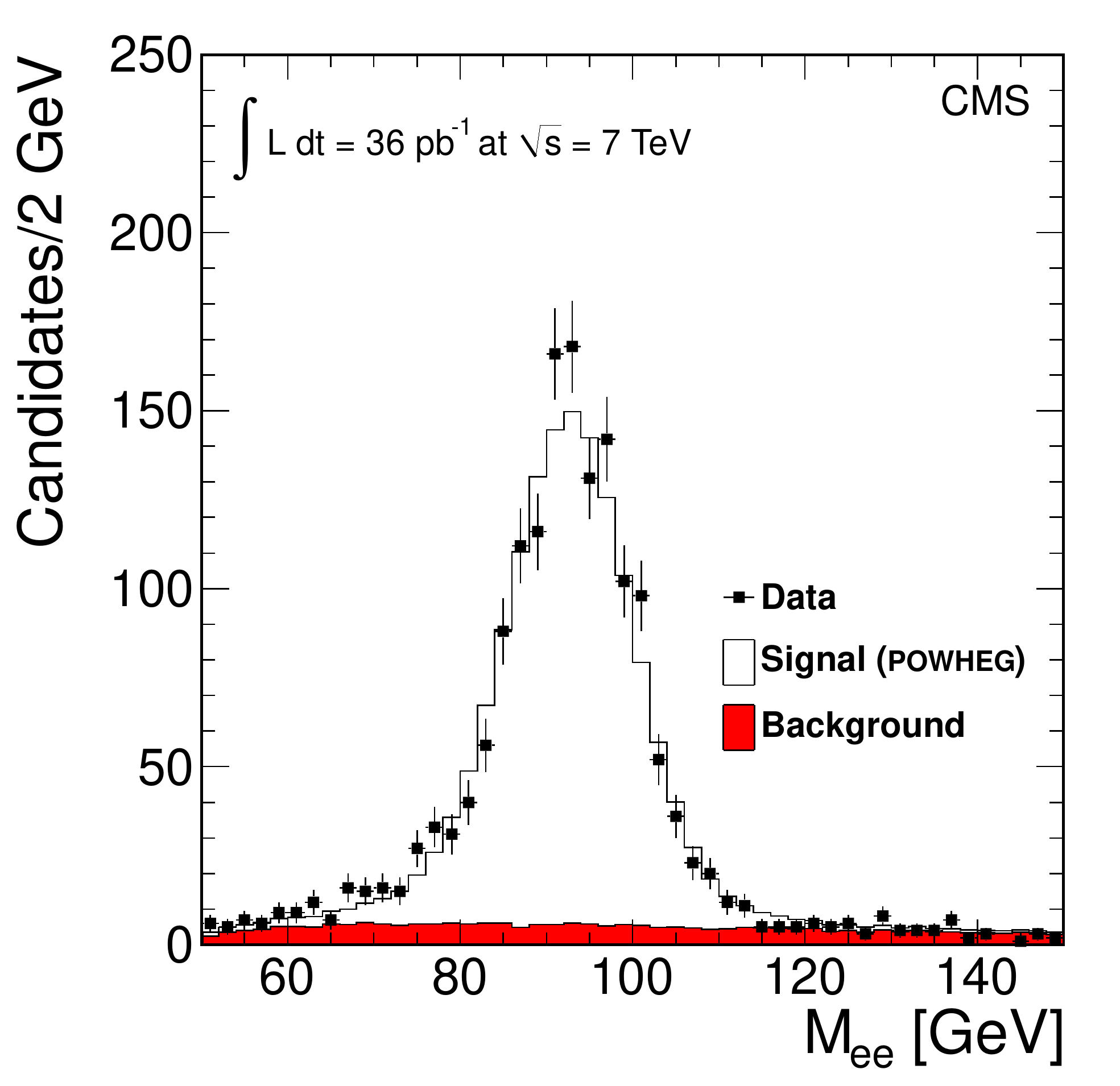}
\end{center}
\caption{\figCaption
  \label{fig:massplots}
}
\end{figure}

Electrons are detected in either the ECAL or the HF.  For this
analysis, the acceptance for electrons is defined to be within the
fiducial region of ECAL, which overlaps with the silicon tracker
region, or in the fiducial region of the HF.  Electrons in this
analysis can thus be observed over pseudorapidity ranges of
$|\eta|<1.444$ (ECAL barrel), $1.566 < |\eta|<2.5$ (ECAL endcaps), and
$3.1 < |\eta| < 4.6$ (HF).  The invariant mass distributions for
selected events in the ECAL-ECAL and the ECAL-HF case are shown
separately in Fig.~\ref{fig:massplots}.  Events are selected online by
a trigger requiring a single electron in the ECAL with $\pt \ge
17\GeV$.  The two electron candidates with highest \pt in the event
are used to reconstruct a \cPZ-boson candidate, and at least one electron
must be in the ECAL and have triggered the event.  No requirement is
applied on the charges of the electrons. Electrons are required to
have $\pt \ge 20\GeV$.  Electrons reconstructed in the ECAL must
have a matching track pointing to the reconstructed electromagnetic
cluster and to be isolated and satisfy the general CMS electron
identification criteria as described in Ref.~\cite{CMS-VBTF-1}.

Electrons are reconstructed in the HF calorimeters from clusters of
3-by-3 towers centered on a seed tower with $\pt > 5\GeV$.  Each tower
provides both a measurement of total energy deposited and the
energy deposited after the 12.5 radiation lengths (22~cm) of absorber
closest to the interaction region.  The two measurements are
approximately equal for high-energy hadrons, while for electromagnetic
particles the second measurement is typically a third of the total
measured energy.  Spurious signals from particles which pass directly
through the phototube windows of the HF are rejected by requiring that
the energy be shared among multiple towers.  Electromagnetic clusters
are selected by requiring the energy in the cluster to be at least
94\% of the energy in the 5-by-5-tower region containing the cluster.
A further selection is performed using the ratio of the two energy
measurements and the ratio of the two most energetic towers in the
cluster to the total cluster energy.

The detector acceptance is obtained from the simulation of the
Drell--Yan process generated with the \POWHEG~\cite{POWHEG}
matrix-element NLO generator interfaced with the \PYTHIA
(v.~6.422)~\cite{PYTHIA} parton-shower event generator, using the CT10
parametrization of the PDFs~\cite{CT10} and the Z2 underlying event
tune~\cite{Field:2010bc}.  The Z2 tune is the standard for CMS
simulation and was tuned to the observed minimum-bias and underlying
event characteristics at $\sqrt{s}=7\TeV$~\cite{UE}.  The effect of
FSR is simulated using \PYTHIA.  In the muon channel, acceptance and
efficiency calculations for the signal are performed using the full
\GEANTfour-based~\cite{GEANT4} detector simulation, with additional
smearing added to correct for observed differences in resolution
between data and simulation as discussed below.  For the electron
channel, a parametrized simulation, matched to the resolution of the
detector as measured in data, was used for efficiency and acceptance
calculations.  For the \qt measurement, the electron acceptance is
restricted to $|\eta|<2.1$ to match the muon acceptance.

The individual lepton detection and selection efficiencies are
determined using a ``tag-and-probe'' method on the candidate lepton
pairs. One of the leptons of the pair, the ``tag'', is required to
pass all the selection requirements. The other lepton, the ``probe'',
is selected with all requirements in the selection up to but excluding
the requirement under study.  The lepton pair is required to have an
invariant mass consistent with the $\cPZ$ boson.  When multiple tag-probe
combinations are possible in a given event, one is chosen at random.
The fraction of the probe leptons that also meet the requirement under
study determines the efficiency of the requirement, after
subtraction of the background from both samples using a fit to the
dilepton invariant mass.  In this manner, the efficiencies for the
reconstruction, isolation, and trigger are measured sequentially.
These efficiencies are compared with the efficiencies determined from
the simulation to produce correction factors, some of which
depend on the lepton kinematics, as discussed below.  The efficiencies
for an electron to form a cluster and a muon to form a basic track,
both of which are very high, are taken from the \GEANTfour
simulation, which includes a modeling of inactive detector regions.
The product of efficiency and acceptance for a given bin of $y$ or \qt
is determined using Monte Carlo simulation as the ratio of the number
of generated events reconstructed in the bin to the number of
generated events before the FSR correction, using the single-lepton
efficiencies determined from data.

The single-muon trigger efficiency is determined separately for the
different data-taking periods and varies from $0.880\pm0.008$ at the
beginning of the period to $0.924\pm0.003$ at the end.  The
single-muon trigger efficiencies are shown to be independent of $\pt$
and $\eta$ within the acceptance used in this analysis. The trigger
efficiency for events with two muons of $\pt>20\GeV$ and $|\eta|<2.1$
is $0.993\pm0.005$, averaged over data-taking periods.  The average
muon reconstruction and identification efficiency for the selection
used in this analysis is $0.950\pm0.003$.  The uncertainty on the
efficiencies is dominated by the data sample size for the
tag-and-probe measurement.

The single-electron trigger efficiency is measured to be between
$0.96\pm0.03$ and $0.99\pm0.01$, varying as a function of $\pt$ and
$\eta$.  For events with both electrons in the ECAL, the event trigger
efficiency is greater than $0.999$.  For the electron channel, the
total reconstruction and identification efficiencies determined from
data range between $0.50$ and $0.90$ and are applied to the simulation
as functions of \pt and $\eta$.  Typical reconstruction and
identification efficiency uncertainties are between 1 and 10\%.
However, the impact of these uncertainties on the final measurement
uncertainty is greatly reduced by the normalization to the total cross
section.

The main sources of background in the measurement are $\cPZ\rightarrow
\tau\tau$ and QCD multijet, \ttbar, $\PW$ +
jets, and diboson production.  Diboson production including a $\cPZ$ is
considered to be a background for the measurement.  All backgrounds
except for QCD multijet production are evaluated using Monte Carlo
simulation.  The $\cPZ\rightarrow \tau\tau$ events are generated with
\POWHEGPYTHIA.  Events from $\ttbar$, diboson
production, and $\PW$ + jets are generated using the \MADGRAPH
(v.~4.4.12)~\cite{MADGRAPH} matrix-element generator interfaced to
\PYTHIA.  Generated events are processed through the full
\GEANTfour-based detector simulation, trigger emulation, and event
reconstruction chain.  We validated the use of the simulation to
determine the background from the $\cPZ\to\tau\tau$,
\ttbar, and diboson backgrounds by analyzing the
\qt spectrum for the $\Pe\mu$ pairs.  These background processes are
flavor-symmetric and produce twice as many $\Pe\mu$ pairs as $\Pe\Pe$ or
$\mu\mu$ pairs.  The analysis of this data sample matched the
expectation from simulation.

The QCD background is estimated using collision data samples.  In the
muon channel, the QCD background is estimated using a nonisolated
dimuon sample corrected for the small contributions in the nonisolated
sample from prompt muons such as those from
\ttbar or \cPZ-boson decay.  The estimate is
verified using a like-sign dimuon sample, since nonprompt sources of
dimuons should have equal rates of like-sign and opposite-sign events.
The QCD background in the muon channel is found to be very small.  In
the electron channel, the QCD background is larger and can be directly
estimated by fitting the dielectron mass distributions in the data.
The fit was performed in each measurement bin over the range $40 <
M_{\Pe\Pe} < 140\GeV$ using a linear combination of a signal shape
from simulation and a background shape determined by inverting
the isolation and electron identification requirements in the data
selection.

After applying all analysis selection criteria, the total background
fraction in the muon channel is $0.4\pm0.4$\%, consisting primarily of
$\cPZ\rightarrow \tau\tau$ and \ttbar
processes and with an uncertainty dominated by statistical
uncertainties in the background simulation.  In the electron channel,
the background fraction is $1.0\pm0.5$\% for $\cPZ$ bosons reconstructed
using two electrons in ECAL and $10\pm4$\% for $\cPZ$ bosons reconstructed
using one electron in ECAL and one in the HF, where the uncertainty is
dominated by statistical uncertainties in the QCD estimate.  In the
electron channel, the QCD background is the dominant background
component in every bin of rapidity and also at low \qt.  In the
highest four \qt bins, the $\cPZ\rightarrow \tau\tau$ and
\ttbar processes are the dominant
contributions to the background.

The bin width in rapidity ($\Delta y = 0.1$) is chosen to allow a comparison
with previous measurements at lower center-of-mass energies.  The bin
widths in \qt, which vary from $2.5$ to $350\GeV$, are chosen to
provide sufficient resolution to observe the shape of the
distribution, to limit bin migration, and to ensure a sufficient data
sample in each measurement bin.

The final measured $y$ and \qt distributions are corrected for
bin-migration effects arising from the detector resolution and from
FSR using a matrix-based unfolding procedure~\cite{bib:unfold2}.
Large simulation samples are used to create the response matrices,
which are inverted and used to unfold the measured distribution.  This
unfolding is applied to allow the combination of the muon and electron
channels, which have different resolutions, and to allow the
comparison with results from other experiments.  The corrections
resulting from detector resolution are calculated by comparing the
generator-level dilepton distribution after FSR obtained from
\POWHEGPYTHIA to that of the reconstructed simulated events, after
smearing the momentum of each lepton with a parametrized function. The
function is derived by comparing the $\cPZ$ mass distribution in
data and the Monte Carlo detector simulation for different regions of
$\eta$ and \pt.  In the muon channel, the smearing represents the
observed difference between the resolution in data and in simulation.  In the
electron channel, as described above, a fully parametrized simulation
is used for the acceptance and efficiency corrections.  The
corrections due to FSR are based on a Monte Carlo simulation using
\PYTHIA, and are obtained by comparing the dilepton $y$ and \qt
distributions before and after FSR.  These corrections are primarily
important for the muon measurements, though large-angle FSR also has
an impact on the electron distributions.

\section{Systematic uncertainties}

The leading sources of systematic uncertainty for the normalized
distribution measurements are the background estimates, the trigger
and identification efficiencies, the unfolding procedure, the
calorimeter energy scale, and the tracker misalignment.  Since the
measurements are normalized by the total measured cross sections,
several sources of systematic uncertainty cancel, as they affect both
the total rate and the differential rate in the same manner.  For
example, the uncertainty in the luminosity measurement cancels
completely and uncertainties resulting from the lepton efficiencies
and from the PDFs are significantly reduced.

The muon and electron measurements share several theory-dependent
uncertainties.  Given the importance of the rapidity measurement in
constraining PDFs, it is crucial to estimate the effect of the
uncertainties in the PDFs on the determination of the bin-by-bin
acceptance for the measurement.  To evaluate the effect, we use the
variations provided as part of the CT10 PDF set~\cite{CT10}.  For this
PDF set, 52 variations are provided, each of which represents a shift
in the PDFs by plus or minus one standard deviation along one of the
26 eigenvectors of the model.  These eigenvectors are used to
parametrize the uncertainties of the PDFs by diagonalizing the actual
PDF model fit parameters, taking into account the unitary requirement
and other constraints.  The eigenvectors are not simply connected to
specific observables, but represent an orthogonal basis in the PDF
model space along which the uncertainties can be calculated.  For each
variation, the effect on the bin-by-bin acceptance normalized by the
total acceptance was determined.  The effects are combined in
quadrature for each bin, separating negative and positive effects, to
give the total uncertainty.  The resulting uncertainties in the
acceptance are less than 0.2\% over the entire measurement range.
However, the change in the shape of the distributions as a function of
$y$ is quite significant, up to 4\% at high rapidity for some
variations.  These changes do not represent systematic uncertainties
in the measurement -- instead they represent the sensitivity of the
analysis for constraining the PDFs.

Several background processes, as described in
Section~\ref{sec:procedure}, are predicted from Monte Carlo simulation
and compared with the data. A conservative estimate of the possible
impact on the measurement is derived by varying the estimates of the
small background from these sources by $100\%$ based on the
uncertainty due to the limited simulation sample size.  We calculate
the deviation of the central value of the normalized distribution in
each bin when the background levels are varied.
For the electron channel, the estimation of the bin-by-bin QCD
background from data is a leading source of systematic uncertainty.
Here, the error on the level of background in the signal region, $60 <
M_{\Pe\Pe} < 120\GeV$, is dominated by the lack of data available
in the dilepton invariant mass sideband regions.

The trigger and the identification efficiencies are measured in the
data as described above.  The largest uncertainty in the efficiencies
arises from the size of the data sample.  To estimate the impact of
these uncertainties on the final measurement, we change the
efficiencies by plus or minus the amount of their statistical
uncertainties and determine the changes of the normalized
distribution.  The changes from the central value are assigned as the
systematic uncertainty arising from the efficiency measurements,
taking into account the cancellation effect from the rate
normalization. The efficiencies from each stage of the selection are
considered independently, and the resulting uncertainties are summed
in quadrature.

The systematic uncertainty from the unfolding procedure is estimated
using alternative response matrices derived in several ways.  We
consider different generator models for the \qt spectrum, which can
affect the distribution of events within the bins.  We vary the
parameters of the detector resolution functions within their
uncertainties.  We also reweight the smeared spectrum to match the
data and evaluate the differences between the nominal and the
reweighted unfolded spectra.  In all cases, the effects amount to
less than 0.5\%.

For the electron channel, the imperfect knowledge of the absolute and
relative energy scales in the electromagnetic and forward calorimeters
is a source of systematic uncertainty.  Using Monte Carlo simulations,
we estimate the effect of the scale uncertainties by scaling the
energies of electrons by amounts corresponding to the calibration
uncertainties and the difference observed between the different
calibration techniques used in the calorimeters.  These energy scale
uncertainties depend on the position of the electron within the
calorimeters.  We then determine the impact of these shifts on the
observed distributions.

\begin{table*}[tbp]
\caption{
Fractional systematic uncertainty contributions for representative rapidity bins
and transverse momentum bins in the electron and muon channels.
\label{tab:syst}
}
\begin{center}

\begin{tabular}{|l|ccccc|} \hline
$|y|$ Range              & \multicolumn{2}{c}{$[0.0,0.1]$} & \multicolumn{2}{c}{$[1.8,1.9]$} & $[3.0,3.1]$ \\
Channel                     & Muon & Electron & Muon & Electron & Electron \\ \hline
Background Estimation       &     0.002       &  0.010        &   0.002        &   0.015        &    0.047      \\
Efficiency Determination    &     0.003       &  0.005        &   0.007        &   0.007        &    0.047      \\
Energy/Momentum Scale       &     0.001       &  0.004        &   0.001        &   0.003        &    0.009      \\
PDF Acceptance Determination&     0.001       &  0.001        &   0.001        &   0.001        &    0.001      \\ \hline
Total                       &     0.004       &  0.012        &   0.007        &   0.017        &    0.067      \\
\hline
\end{tabular}

\vspace{1ex}

\begin{tabular}{|l|cccc|} \hline
\qt Range                   & \multicolumn{2}{c}{$[2.5\GeV,5.0\GeV]$} & \multicolumn{2}{c|}{$[110\GeV,150\GeV]$} \\
Channel                     & Muon & Electron & Muon & Electron \\ \hline
Background Estimation       &       0.004       & 0.005        &   0.019  &   0.028               \\
Efficiency Determination    &       0.010       & 0.002        &   0.010  &   0.008               \\
Energy Scale                &        --         & 0.022        &    --    &   0.035               \\
Tracker Alignment           &       0.015       & 0.013        &   0.023  &   0.020               \\
Unfolding                   &       0.006       & 0.004        &   0.017  &   0.001               \\
PDF Acceptance Determination &      0.002       & 0.002        &   0.001  &   0.001               \\ \hline
Total                       &       0.020       & 0.026        &   0.036  &   0.050 \\
\hline
\end{tabular}

\end{center}
\end{table*}

The muon \pt used in the analysis is based on the silicon tracker
measurement. Thus, any misalignment of the tracker may directly affect
the muon momentum resolution. The systematic uncertainty associated
with tracker misalignment is calculated by reprocessing the Drell--Yan
simulation using several models designed to reproduce the possible
misalignments that may be present in the tracker. The bin-by-bin
maximum deviation from the nominal Drell--Yan simulation is used as
estimator of the tracker misalignment uncertainty.  In the electron
channel, the sensitivity to the tracker alignment is determined by
comparing the reconstructed $y$ and \pt using the calorimeter energy
alone with those including the track measurements, for both data and
simulation.

The systematic uncertainties are summarized in Table~\ref{tab:syst} for
representative values of $y$ and \qt in the muon and electron channels.
After combining the effects discussed above, the total systematic
uncertainty in each bin is found to be significantly smaller than the statistical
uncertainty.

\section{Rapidity Distribution Results}

The rapidity $y$ of $\cPZ$ bosons produced in proton-proton collisions is
related to the momentum fraction $x_+$ ($x_-$) carried by the parton
in the forward-going (backward-going) proton as described by the
leading-order formula \( x_\pm=\frac{m_\cPZ}{\sqrt{s}} e^{\pm y} \).
Therefore, the rapidity distribution directly reflects the PDFs of the
interacting partons.  At the LHC, the rapidity distribution of $\cPZ$ bosons is expected to be symmetric around zero, therefore the
appropriate measurement is the distribution of $\cPZ$ bosons as a function
of the absolute value of rapidity.  The measurement is normalized to
the total cross section ($1/\sigma\; d\sigma/d\left|y\right|$), where
$\sigma$ is the cross section determined by the sum of all
observed $y$ bins ($|y|<3.5$), corrected to the total cross section as
calculated from \POWHEG with CT10 PDFs.  The calculated correction
between the measured and total $y$ range is 0.983 with an uncertainty
of 0.001 from PDF variation.

\begin{figure}[tbph]
\begin{center}
\includegraphics[width=\cmsLargefigwid]{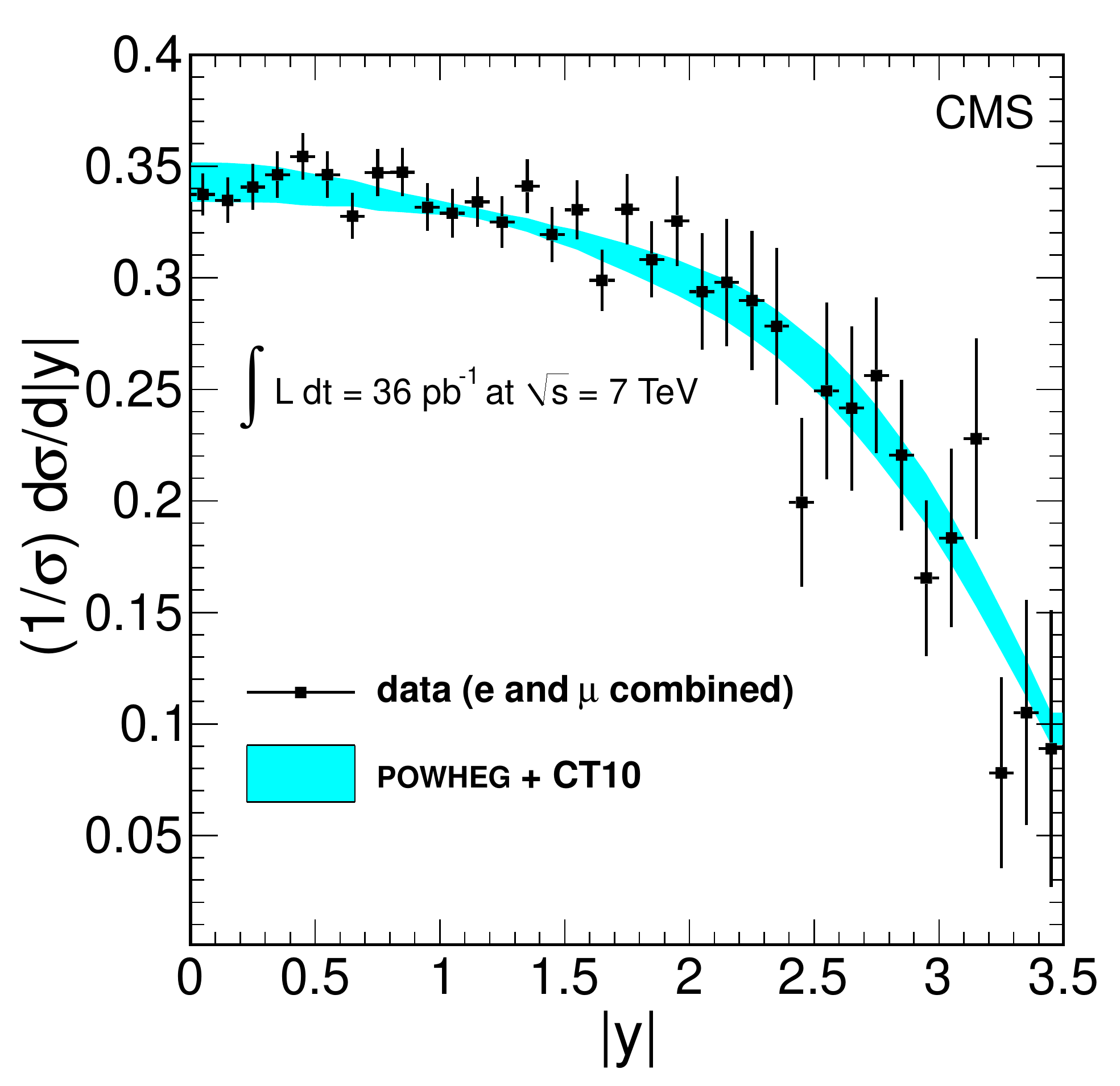}
\caption{ The normalized differential cross section for $\cPZ$ bosons as a function of
  the absolute value of rapidity, combining the muon and
  electron channels.  The error bars correspond to the experimental
  statistical and systematic uncertainties added in quadrature.  The
  shaded area indicates the range of variation predicted by the \POWHEG
  simulation for the uncertainties of the CT10 PDFs.  }
\label{fig:combo_y}
\end{center}
\end{figure}

The measurements for the muon and electron channels are given in
Table~\ref{tab:combo_y} and are in agreement with each other
(reduced $\chi^2=0.85$) over the 20 bins where the measurements
overlap.  We combine these two measurements using the procedure
defined in Ref.~\cite{BLUE}, which provides a full covariance matrix
for the uncertainties.  The uncertainties are considered to be uncorrelated
between the two analyses, since the only correlation between the
channels is from the  small PDF uncertainty.  The combined measurements
are shown in Table~\ref{tab:combo_y} and compared to the predictions
made using CT10 PDFs in Fig.~\ref{fig:combo_y}.

\begin{table}[tbp]
\caption{
Measurement of the normalized differential cross section $\left(\frac{1}{\sigma}\frac{\mathrm{d}\sigma}{\mathrm{d}|y|}\right)$
for Drell--Yan lepton pairs in the \cPZ-boson mass region ($60 < M_{\ell\ell} < 120\GeV$) as a function of
the absolute value of rapidity, separately for the muon and electron channels and combined.
Detector geometry and trigger uniformity requirements limit the muon channel measurement to $|y|<2.0$.
The uncertainties shown are the combined
statistical and systematic uncertainties.
\label{tab:combo_y}
}
\begin{center}
\begin{tabular}{|c||l@{$\pm$}l|l@{$\pm$}l|l@{$\pm$}l|}
\hline
   & \multicolumn{6}{c|}{Normalized Differential Cross section} \\
$|y|$ Range & \multicolumn{2}{c|}{Muon} & \multicolumn{2}{c|}{Electron} & \multicolumn{2}{c|}{Combined}   \\ \hline
  $[ 0.0 ,     0.1]$ &      0.324 &    0.012 &  0.359 &    0.015 & 0.337 &    0.010 \\
  $[ 0.1 ,     0.2]$ &      0.338 &    0.013 &  0.326 &    0.016 & 0.335 &    0.010 \\
  $[ 0.2 ,     0.3]$ &      0.338 &    0.013 &  0.344 &    0.017 & 0.341 &    0.010 \\
  $[ 0.3 ,     0.4]$ &      0.341 &    0.013 &  0.355 &    0.017 & 0.346 &    0.010 \\
  $[ 0.4 ,     0.5]$ &      0.363 &    0.013 &  0.339 &    0.017 & 0.354 &    0.011 \\
  $[ 0.5 ,     0.6]$ &      0.342 &    0.013 &  0.351 &    0.018 & 0.346 &    0.010 \\
  $[ 0.6 ,     0.7]$ &      0.312 &    0.013 &  0.360 &    0.018 & 0.328 &    0.010 \\
  $[ 0.7 ,     0.8]$ &      0.354 &    0.013 &  0.331 &    0.018 & 0.347 &    0.011 \\
  $[ 0.8 ,     0.9]$ &      0.343 &    0.014 &  0.355 &    0.018 & 0.347 &    0.011 \\
  $[ 0.9 ,     1.0]$ &      0.332 &    0.014 &  0.332 &    0.018 & 0.332 &    0.011 \\
  $[ 1.0 ,     1.1]$ &      0.336 &    0.014 &  0.316 &    0.018 & 0.329 &    0.011 \\
  $[ 1.1 ,     1.2]$ &      0.324 &    0.014 &  0.352 &    0.019 & 0.334 &    0.011 \\
  $[ 1.2 ,     1.3]$ &      0.321 &    0.014 &  0.332 &    0.019 & 0.325 &    0.011 \\
  $[ 1.3 ,     1.4]$ &      0.355 &    0.016 &  0.321 &    0.019 & 0.341 &    0.012 \\
  $[ 1.4 ,     1.5]$ &      0.326 &    0.016 &  0.313 &    0.019 & 0.319 &    0.012 \\
  $[ 1.5 ,     1.6]$ &      0.331 &    0.018 &  0.330 &    0.020 & 0.330 &    0.013 \\
  $[ 1.6 ,     1.7]$ &      0.294 &    0.018 &  0.306 &    0.022 & 0.299 &    0.014 \\
  $[ 1.7 ,     1.8]$ &      0.331 &    0.021 &  0.332 &    0.024 & 0.331 &    0.016 \\
  $[ 1.8 ,     1.9]$ &      0.324 &    0.025 &  0.294 &    0.024 & 0.308 &    0.017 \\
  $[ 1.9 ,     2.0]$ &      0.328 &    0.032 &  0.328 &    0.026 & 0.328 &    0.020 \\
  $[ 2.0 ,     2.1]$ &    \multicolumn{2}{c|}{ } & 0.294 &    0.027 &  0.294 &    0.027 \\
  $[ 2.1 ,     2.2]$ &    \multicolumn{2}{c|}{ } & 0.298 &    0.029 &  0.298 &    0.029 \\
  $[ 2.2 ,     2.3]$ &    \multicolumn{2}{c|}{ } & 0.290 &    0.031 &  0.290 &    0.031 \\
  $[ 2.3 ,     2.4]$ &    \multicolumn{2}{c|}{ } & 0.278 &    0.035 &  0.278 &    0.035 \\
  $[ 2.4 ,     2.5]$ &    \multicolumn{2}{c|}{ } & 0.199 &    0.038 &  0.199 &    0.038 \\
  $[ 2.5 ,     2.6]$ &    \multicolumn{2}{c|}{ } & 0.249 &    0.040 &  0.249 &    0.040 \\
  $[ 2.6 ,     2.7]$ &    \multicolumn{2}{c|}{ } & 0.241 &    0.037 &  0.241 &    0.037 \\
  $[ 2.7 ,     2.8]$ &    \multicolumn{2}{c|}{ } & 0.256 &    0.035 &  0.256 &    0.035 \\
  $[ 2.8 ,     2.9]$ &    \multicolumn{2}{c|}{ } & 0.221 &    0.034 &  0.221 &    0.034 \\
  $[ 2.9 ,     3.0]$ &    \multicolumn{2}{c|}{ } & 0.165 &    0.035 &  0.165 &    0.035 \\
  $[ 3.0 ,     3.1]$ &    \multicolumn{2}{c|}{ } & 0.183 &    0.040 &  0.183 &    0.040 \\
  $[ 3.1 ,     3.2]$ &    \multicolumn{2}{c|}{ } & 0.228 &    0.045 &  0.228 &    0.045 \\
  $[ 3.2 ,     3.3]$ &    \multicolumn{2}{c|}{ } & 0.078 &    0.043 &  0.078 &    0.043 \\
  $[ 3.3 ,     3.4]$ &    \multicolumn{2}{c|}{ } & 0.105 &    0.051 &  0.105 &    0.051 \\
  $[ 3.4 ,     3.5]$ &    \multicolumn{2}{c|}{ } & 0.089 &    0.062 &  0.089 &    0.062 \\
\hline
\end{tabular}

\end{center}
\end{table}

\begin{figure}[phtb]
\begin{center}
\includegraphics[width=\cmsLargefigwid]{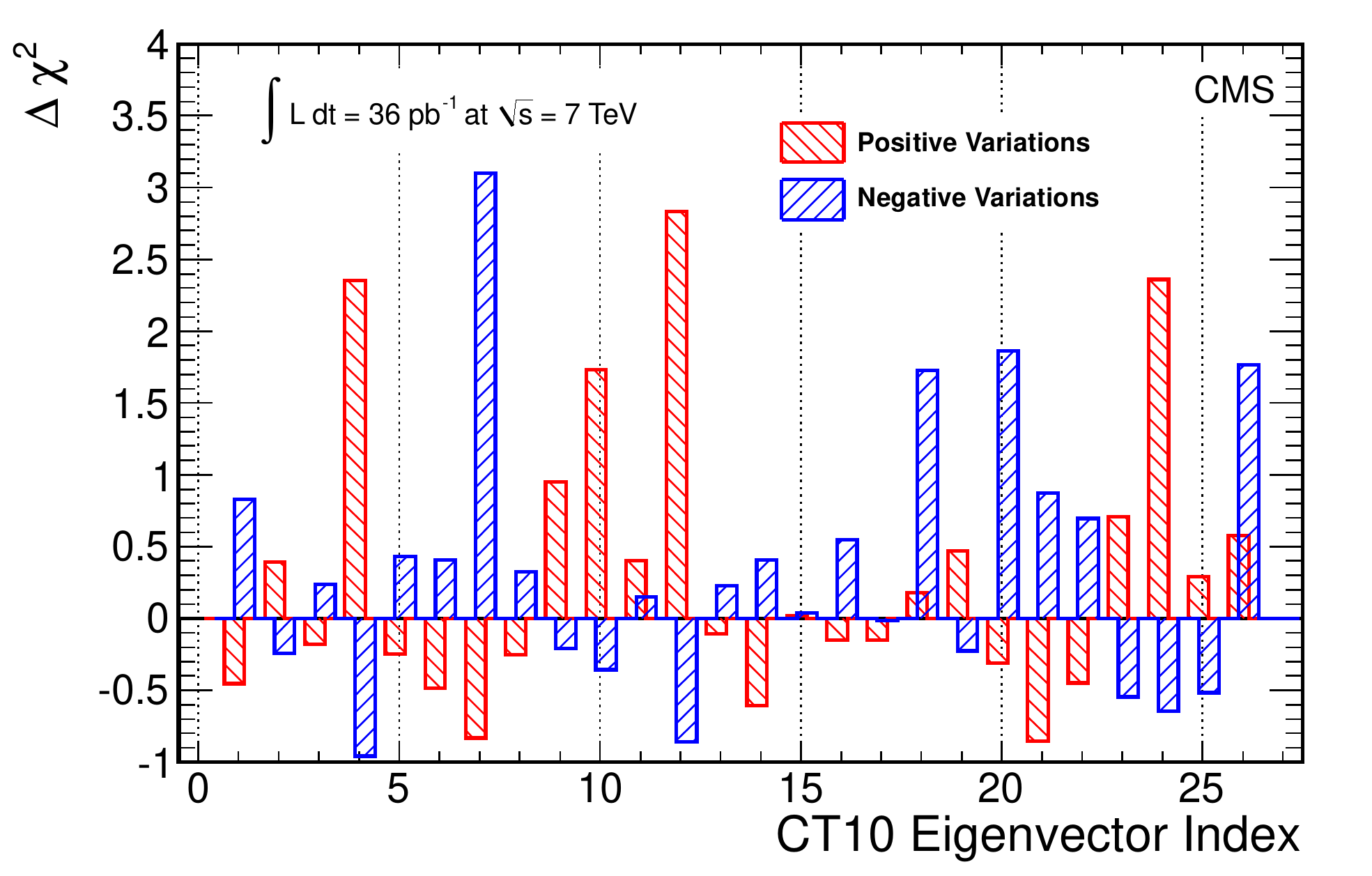}
\end{center}
\caption{The change in $\chi^2$ when comparing the $\cPZ$ rapidity
  differential cross section measurement with the predictions of the
  NLO CT10 PDF set as each of the eigenvector input parameters is
  varied by plus or minus one standard deviation around its default
  value.\label{fig:pdf_chi2_ct10}}
\end{figure}

\begin{figure}[phtb]
\begin{center}
\includegraphics[width=\cmsLargefigwid]{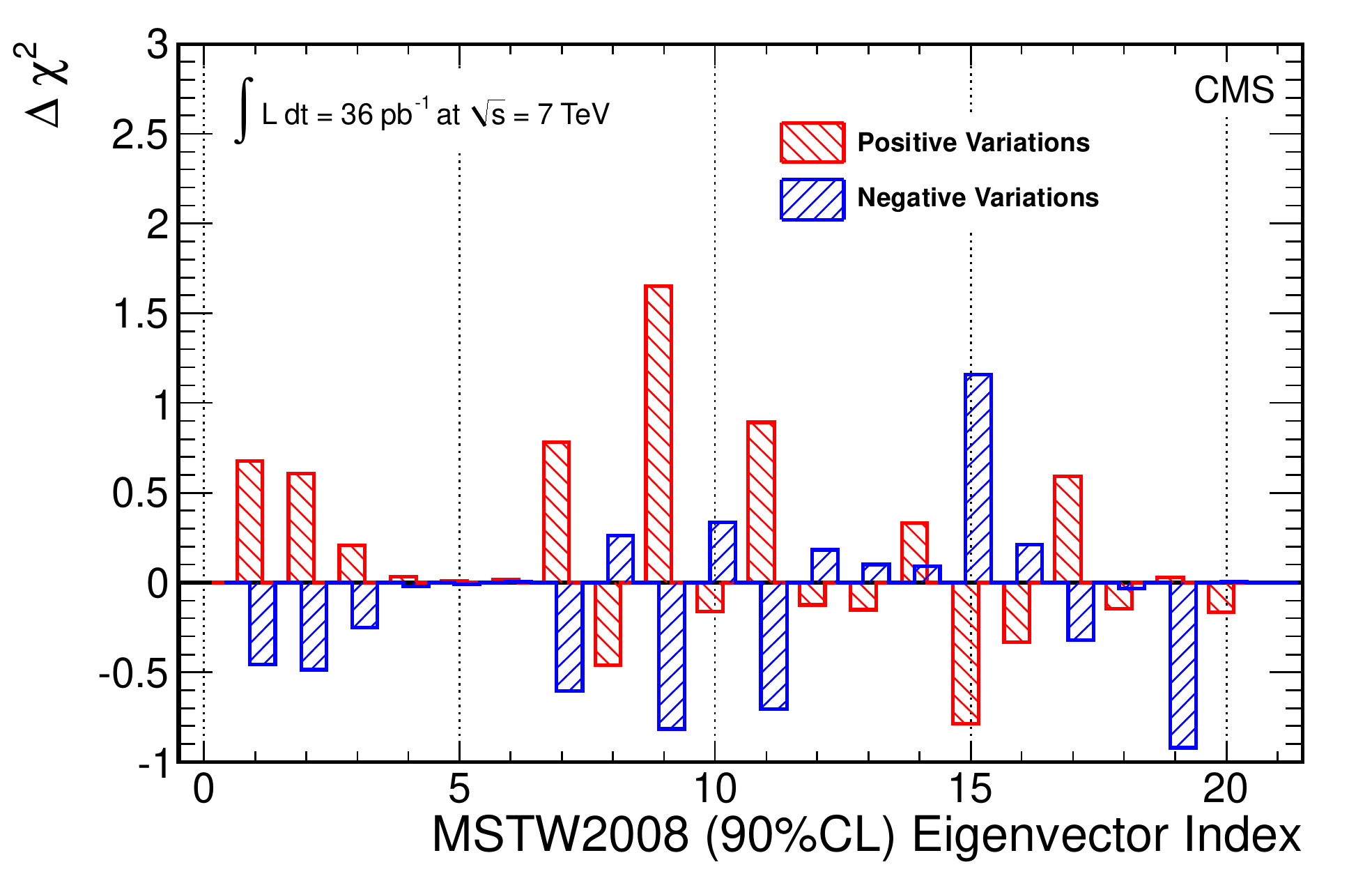}
\end{center}
\caption{The change in $\chi^2$ when comparing the $\cPZ$ rapidity
  differential cross section measurement with the predictions
 of the NLO MST2008 PDF set as each of the eigenvector input parameters is varied by $\pm 90$\% confidence level (CL)
 around its default value.
 \label{fig:pdf_chi2_mstw08}}
\end{figure}

\begin{figure}[phtb]
\begin{center}
\includegraphics[width=\cmsLargefigwid]{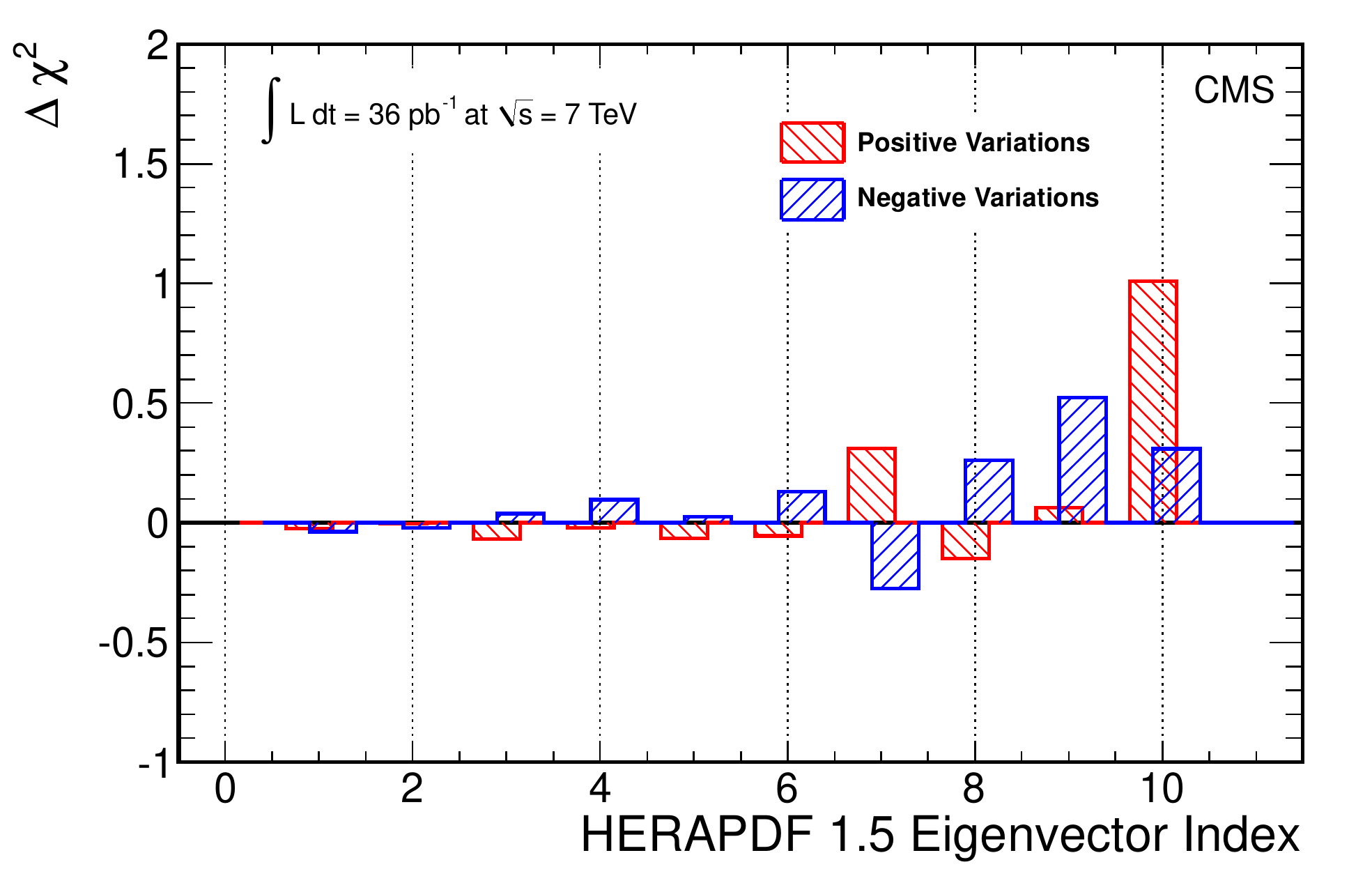}
\includegraphics[width=\cmsLargefigwid]{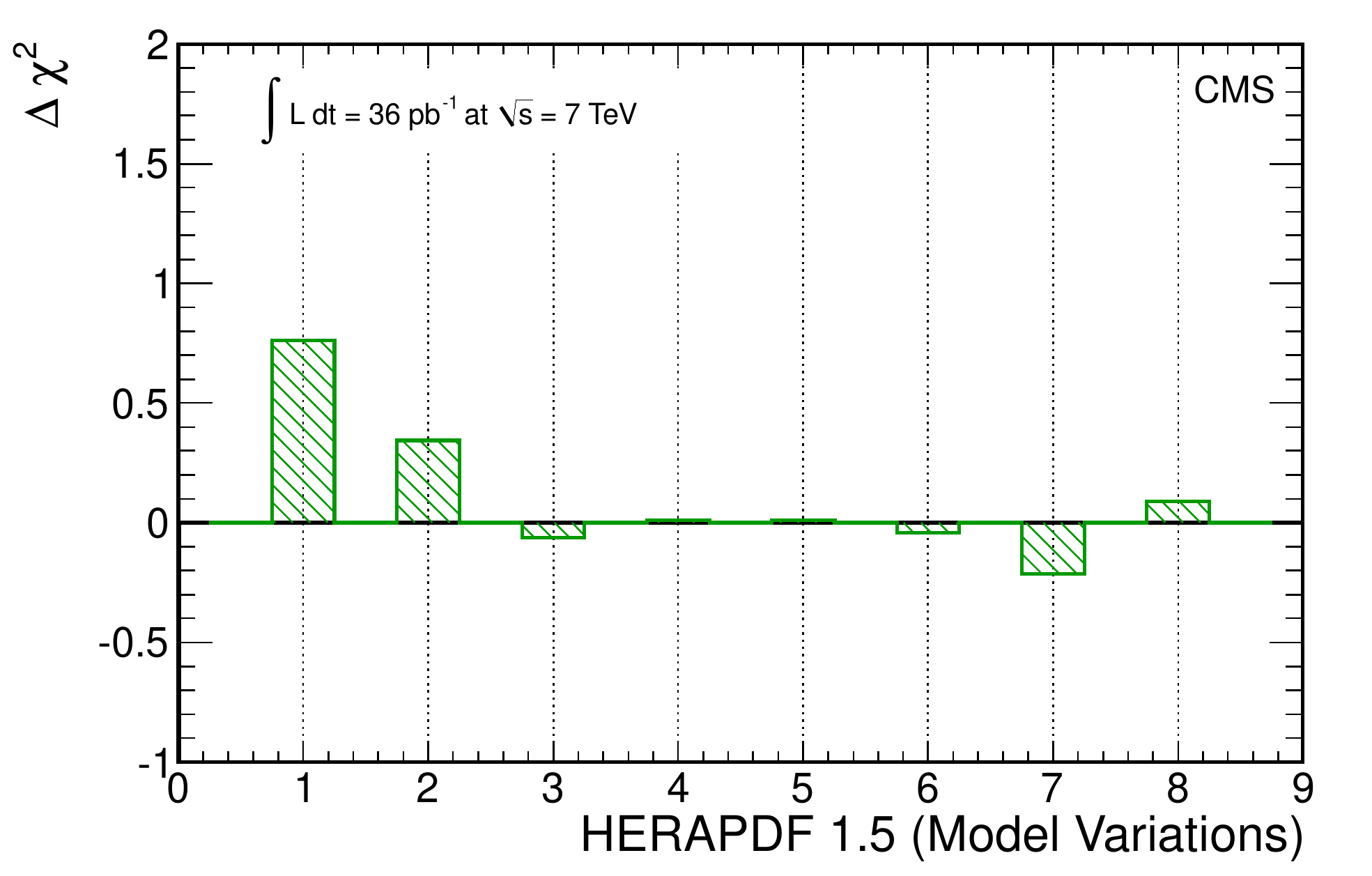}
\end{center}
\caption{The change in $\chi^2$ when comparing the $\cPZ$ rapidity
  differential cross section measurement with the predictions
  of the NLO HERAPDF~1.5 PDF set as each of the eigenvector input parameters (\xleft) and
  the model parameters (\xright) is varied by one standard deviation around its default value.
  These together represent the full set of uncertainties in the HERAPDF~1.5 set.
  \label{fig:pdf_chi2_herapdf}}
\end{figure}

To evaluate the sensitivity of this result to parameters of some of
the more-recent PDF sets, we determine the change in the $\chi^2$ for
each variation of the eigenvectors provided in the PDF sets.  The CT10
PDF set has a $\chi^2$ of 18.5 for the base prediction, and the
eigenvector-dependent changes in $\chi^2$ are shown in
Fig.~\ref{fig:pdf_chi2_ct10}.  The number of degrees of freedom (ndof)
is 34.  The MSTW2008~\cite{MSTW2008} PDF set has a $\chi^2$ of 18.3
for its base prediction, and the eigenvector-dependent changes shown
in Fig.~\ref{fig:pdf_chi2_mstw08}.  For both sets, several
eigenvectors show significant sensitivity to our result, with CT10
showing a generally larger sensitivity.  The
HERAPDF~1.5~\cite{HERAPDF15} PDF set, which has a $\chi^2$ of 18.4 for
its base prediction, provides both eigenvectors and model dependencies
as part of the PDF set.  The changes in $\chi^2$ for both are shown in
Fig.~\ref{fig:pdf_chi2_herapdf}.  The largest model dependencies with
our measurement are the strange-quark PDF as a fraction of the
down-quark-sea PDF.  For the NNPDF~2.0 PDF set~\cite{NNPDF20}, the
base prediction has a $\chi^2$ of 18.4. The NNPDF formalism does not
use eigenvectors, but rather replica PDFs sampled from the same space.
In comparing our result with the 100 standard
NNPDF~2.0 replicas, the majority have $\chi^2$ similar to
the base, but some have $\chi^2$ values up to 34.5,
indicating that these replicas are disfavored significantly by the new
measurement.

\section{Transverse Momentum Distribution Results}

Measurements of the \qt distribution for $\cPZ$ bosons provide an important
test of the QCD predictions of the initial-state gluon-radiation
process.  Perturbative QCD calculations are expected to provide a
reliable prediction for the portion of the spectrum $\qt>20\GeV$,
which is dominated by single hard-gluon emission.  For $\qt<10\GeV$,
the shape of the distribution is determined by multiple soft gluon
radiation and nonperturbative effects.  Such effects are simulated by
Monte Carlo programs combining parton showering and parameterized
models.  These soft-gluon contributions can also be accounted for by
resummation calculations in some Monte Carlo
programs.

For the \qt measurement, the data are normalized to the
cross section integrated over the acceptance region $|\eta|<2.1$ and
$\PT>20\GeV$.  The lepton \pt and $|\eta|$ restrictions
apply to both leptons of a dilepton pair.  The restriction on the
electron pseudorapidity (compared to that used for the rapidity
measurement) allows the combination of the two channels and a more
straightforward physics interpretation, as the two measurements refer
to the same rapidity range and have the same PDF dependence.

\begin{table*}[htbp]
  \caption{
    Measurement of the normalized differential cross section for
Drell--Yan lepton pairs in the \cPZ-boson mass region ($60 < M_{\ell\ell} < 120\GeV$)
  as a function of
    \qt, separately for muon and electron channels
    and for the combination of the two channels.
    The distribution is normalized by the cross section for $\cPZ$ bosons with both leptons having
    $|\eta|<2.1$ and $\PT>20\GeV$.
    The uncertainties listed in the table are
    the combined statistical and systematic uncertainties.
  }
\label{tab:combo}
\begin{center}
\begin{tabular}{|c|c|c|c|}
\hline
\qt Range (\GeV)     & Muon Channel                & Electron Channel            & Combination \\ \hline
 $[   0.0,    2.5]$  &$(3.21\pm0.14)\times 10^{-2}$& $(3.24\pm0.25)\times 10^{-2}$& $(3.22\pm0.13)\times 10^{-2}$\\
 $[   2.5,    5.0]$  &$(5.89\pm0.21)\times 10^{-2}$& $(6.03\pm0.32)\times 10^{-2}$& $(5.92\pm0.17)\times 10^{-2}$\\
 $[   5.0,    7.5]$  &$(5.51\pm0.20)\times 10^{-2}$& $(5.32\pm0.32)\times 10^{-2}$& $(5.50\pm0.16)\times 10^{-2}$\\
 $[   7.5,   10.0]$  &$(3.90\pm0.18)\times 10^{-2}$& $(4.20\pm0.30)\times 10^{-2}$& $(3.96\pm0.14)\times 10^{-2}$\\
 $[  10.0,   12.5]$  &$(3.49\pm0.16)\times 10^{-2}$& $(3.60\pm0.28)\times 10^{-2}$& $(3.53\pm0.12)\times 10^{-2}$\\
 $[  12.5,   15.0]$  &$(2.74\pm0.15)\times 10^{-2}$& $(2.70\pm0.25)\times 10^{-2}$& $(2.72\pm0.12)\times 10^{-2}$\\
 $[  15.0,   17.5]$  &$(2.23\pm0.14)\times 10^{-2}$& $(2.00\pm0.22)\times 10^{-2}$& $(2.16\pm0.10)\times 10^{-2}$\\
 $[  17.5,   20.0]$  &$(1.68\pm0.12)\times 10^{-2}$& $(1.59\pm0.20)\times 10^{-2}$& $(1.65\pm0.09)\times 10^{-2}$\\
 $[  20.0,   30.0]$  &$(1.14\pm0.04)\times 10^{-2}$& $(1.20\pm0.05)\times 10^{-2}$& $(1.16\pm0.04)\times 10^{-2}$\\
 $[  30.0,   40.0]$  &$(6.32\pm0.28)\times 10^{-3}$& $(5.62\pm0.31)\times 10^{-3}$& $(5.98\pm0.27)\times 10^{-3}$\\
 $[  40.0,   50.0]$  &$(3.53\pm0.21)\times 10^{-3}$& $(3.18\pm0.24)\times 10^{-3}$& $(3.38\pm0.18)\times 10^{-3}$\\
 $[  50.0,   70.0]$  &$(1.74\pm0.10)\times 10^{-3}$& $(1.90\pm0.12)\times 10^{-3}$& $(1.81\pm0.09)\times 10^{-3}$\\
 $[  70.0,   90.0]$  &$(7.76\pm0.71)\times 10^{-4}$& $(7.86\pm0.77)\times 10^{-4}$& $(7.79\pm0.54)\times 10^{-4}$\\
 $[  90.0,  110.0]$  &$(4.87\pm0.55)\times 10^{-4}$& $(4.57\pm0.59)\times 10^{-4}$& $(4.75\pm0.42)\times 10^{-4}$\\
 $[ 110.0,  150.0]$  &$(1.79\pm0.22)\times 10^{-4}$& $(2.18\pm0.26)\times 10^{-4}$& $(1.93\pm0.17)\times 10^{-4}$\\
 $[ 150.0,  190.0]$  &$(7.10\pm1.40)\times 10^{-5}$& $(4.82\pm1.31)\times 10^{-5}$& $(6.00\pm0.99)\times 10^{-5}$\\
 $[ 190.0,  250.0]$  &$(1.17\pm0.51)\times 10^{-5}$& $(2.05\pm0.64)\times 10^{-5}$& $(1.51\pm0.43)\times 10^{-5}$\\
 $[ 250.0,  600.0]$  &$(2.24\pm0.78)\times 10^{-6}$& $(0.81\pm0.52)\times 10^{-6}$& $(1.29\pm0.44)\times 10^{-6}$\\
\hline
\end{tabular}

\end{center}
\end{table*}

\begin{figure}[tbph]
\begin{center}
\includegraphics[width=\cmsLargefigwid]{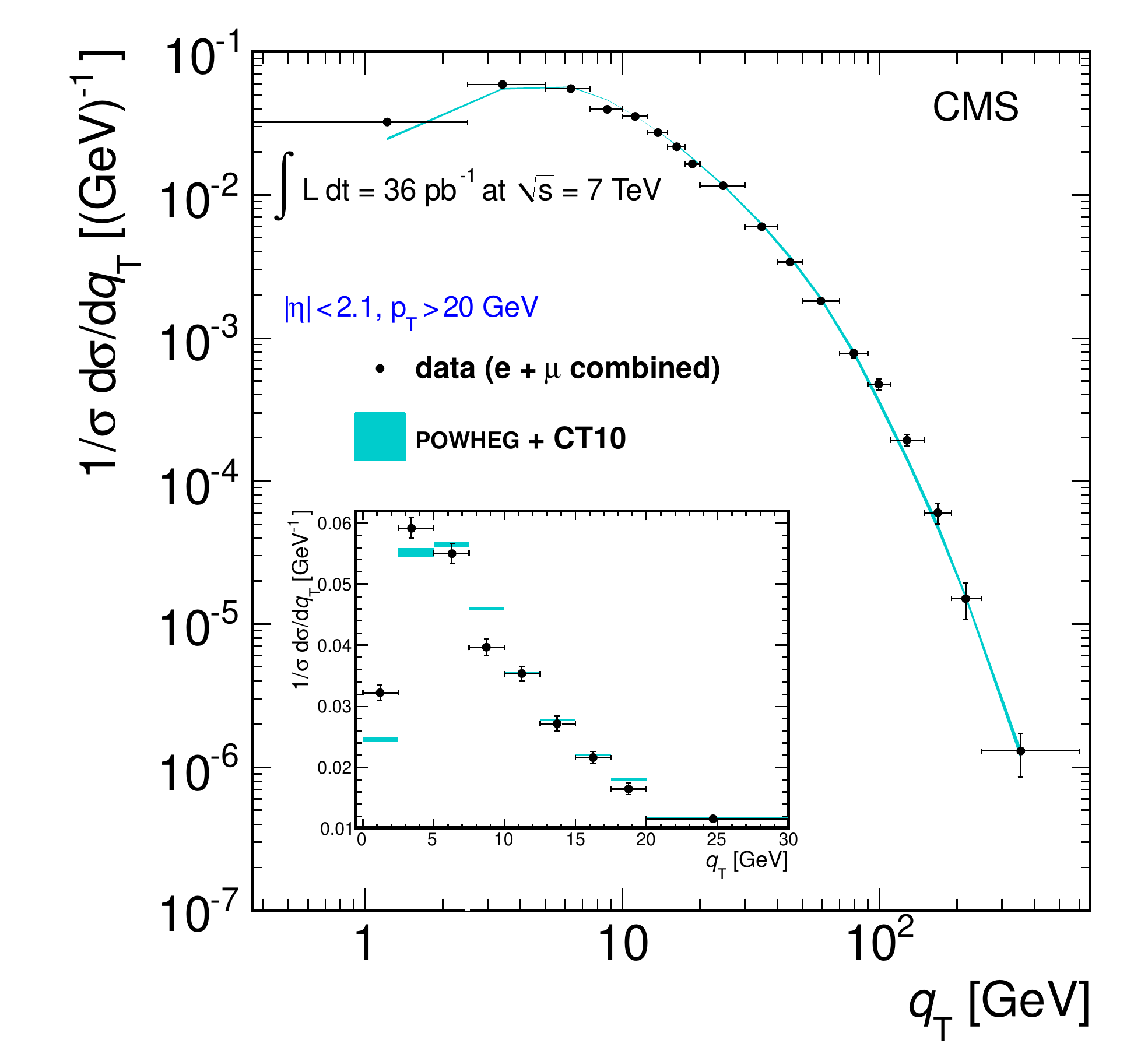}
\end{center}
\caption{ The \cPZ-boson transverse momentum distribution found from
  combining the muon and electron channels, compared to the
  predictions of the \POWHEG generator interfaced with \PYTHIA using
  the Z2 tune.  The error bars correspond to the statistical and
  systematic uncertainties added in quadrature. The band around the
  theoretical prediction includes the uncertainties due to scale
  variations and PDFs.  The horizontal error bars indicate the bin
  boundaries, and the data points are positioned at the
  center-of-gravity of the bins, based on the \POWHEG prediction.  The
  inset figure shows the low \qt region on a linear
  scale.}
\label{fig:combo}
\end{figure}

The measurements from the muon and electron channels are tabulated in
Table~\ref{tab:combo} and are found to be compatible with each other
over the full \qt range (reduced $\chi^2=0.74$).  The combination of
the muon and electron results is also performed following
Ref.~\cite{BLUE}.  The alignment uncertainty is treated as correlated
between the two channels, and other uncertainties are treated as
uncorrelated.  The combined measurement is presented in
Fig.~\ref{fig:combo}, where the data points are positioned at the
center-of-gravity of the bins, based on the \POWHEG prediction.  For
$\qt>20~\GeV$, we compare the data and the prediction of \POWHEGPYTHIA
with the Z2 tune and find $\chi^2/\mathrm{ndof}=19.1/9$, where ndof is
equal to the number of points minus one because of the
normalization. We have taken the full covariance matrix into account
when computing the $\chi^2$ values.  At low momentum, there is poor
agreement, suggesting the need for additional tuning of the
combination of \POWHEG and \PYTHIA in this region, where both
contribute to the observed \qt.

\begin{figure}[phtb]
\begin{center}
\includegraphics[width=\cmsLargefigwid]{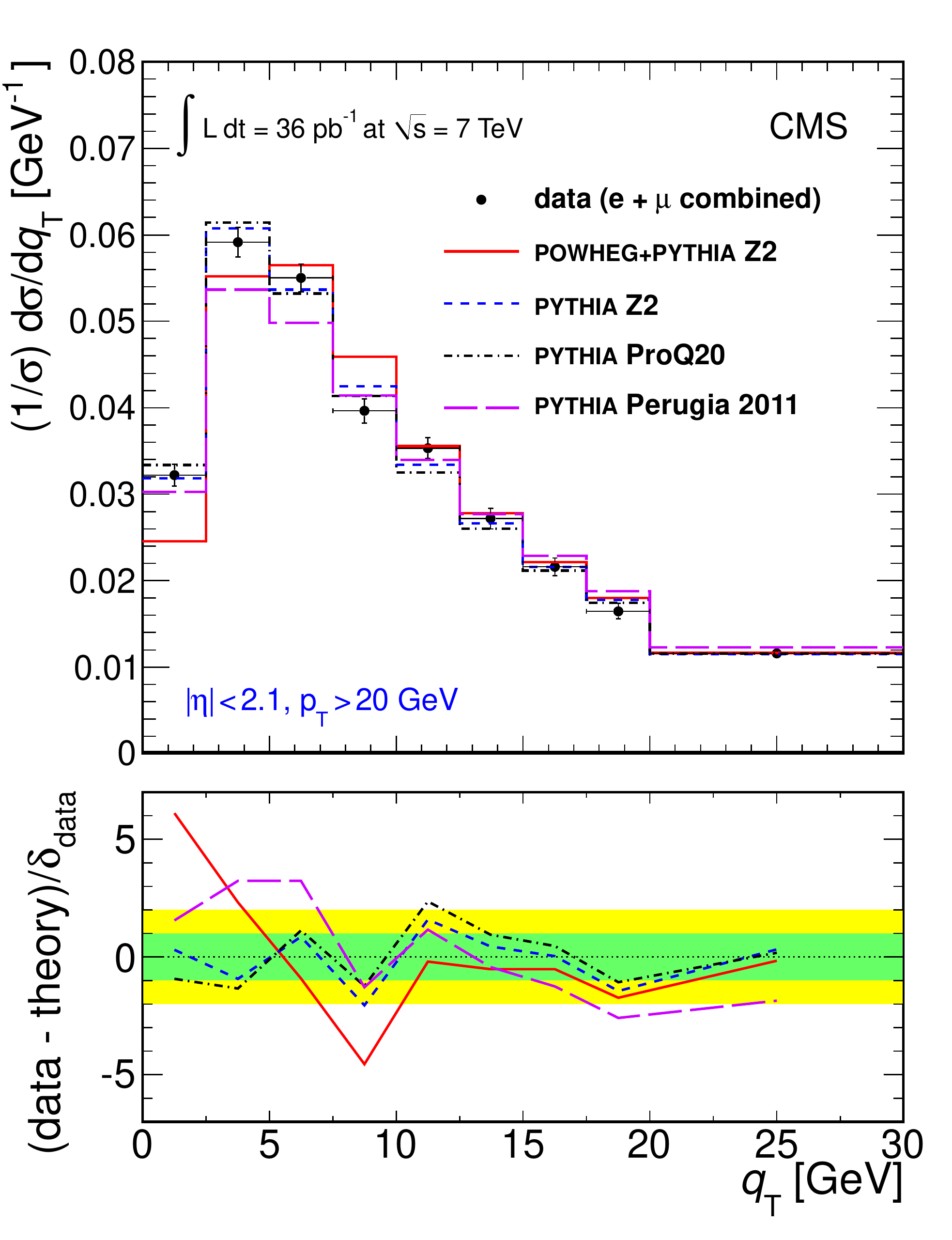}
\caption{ The combined electron and muon measurement of the \cPZ-boson   transverse momentum distribution (points) and the predictions of
  several \PYTHIA tunes and \POWHEG interfaced with \PYTHIA using the
  Z2 tune (histograms).  The error bars on the points represent the
  sum of the statistical and systematic uncertainties on the data.
  The lower portion of the figure shows the difference between the
  data and the simulation predictions divided by the uncertainty
  $\delta$ on the data.  The green (inner) and yellow (outer) bands
  are the $\pm1 \delta$ and $\pm2 \delta$ experimental
  uncertainties. }
\label{fig:combo_pythia}
\end{center}
\end{figure}

At low transverse momenta, {\ie} $\qt < 30\GeV$, the distribution is
determined by nonperturbative QCD, which is modeled by \PYTHIA with a
few free parameters.  Several parameter sets called ``tunes'' are
available, including the Perugia 2011~\cite{Skands:2011PerugiaTunes},
ProQ20~\cite{bib:pro}, and Z2 tune~\cite{Field:2010bc}.  The shapes
predicted with these tunes are compared to this measurement in
Fig.~\ref{fig:combo_pythia}.  Agreement is observed for the Z2
($\chi^2/\mathrm{ndof} = 9.4/8$) and the ProQ20 tunes
($\chi^2/\mathrm{ndof} = 13.3/8$), but disagreement for the Perugia
2011 tune ($\chi^2/\mathrm{ndof} = 48.8/8$) and for \POWHEGPYTHIA
($\chi^2/\mathrm{ndof} = 76.3/8$).  These results provide a validation
of the Z2 tune for a high momentum-scale process that is rather
different from the low-momentum-scale processes that determine the
characteristics of minimum-bias events and the underlying event from
which the parameters of the Z2 tune were originally obtained.

\begin{figure}[hbtp]
  \begin{center}
    \includegraphics[width=\cmsLargefigwid]{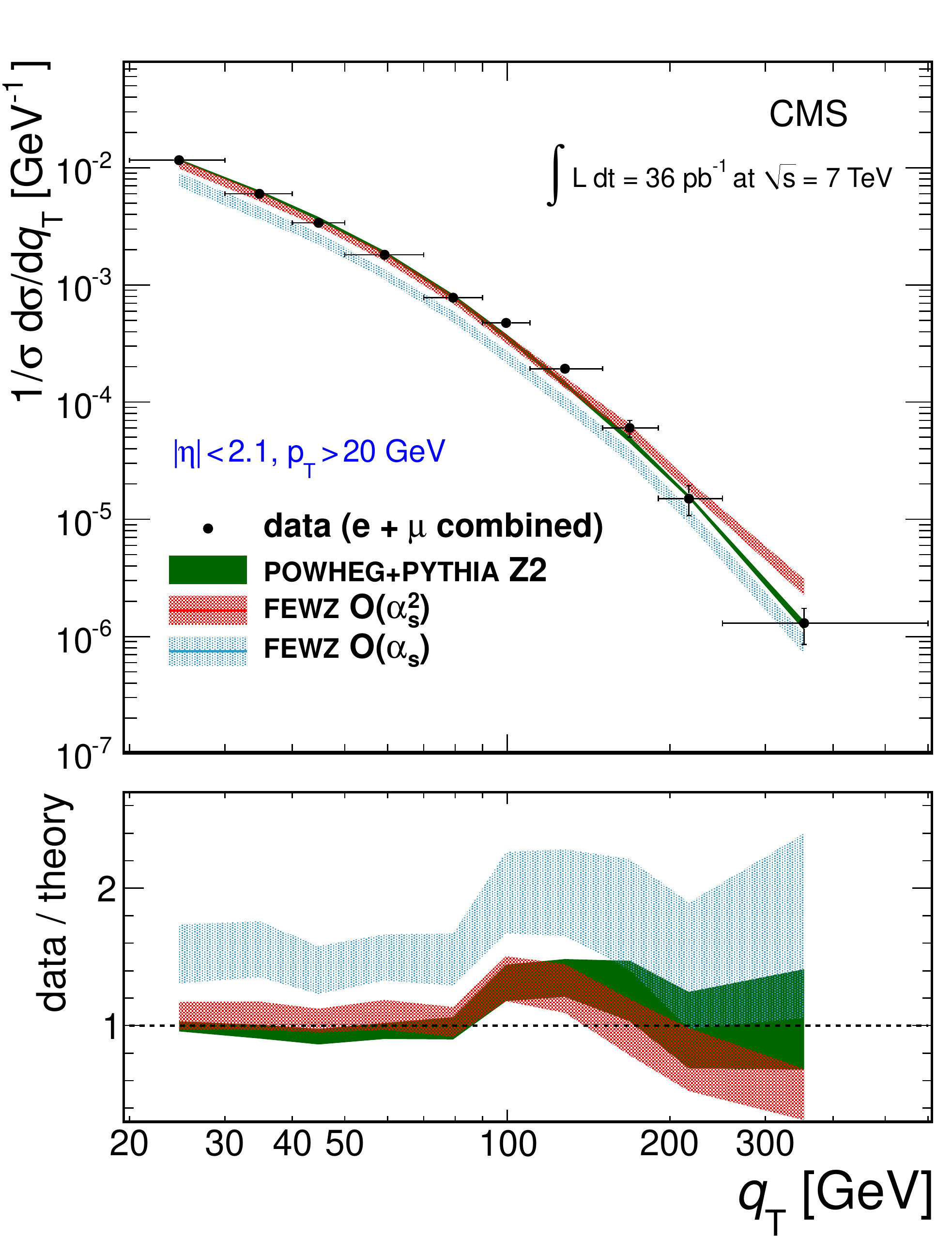}
    \caption{The combined electron and muon \cPZ-boson normalized differential cross section as a function of transverse momentum (points) and the \POWHEG and \FEWZ predictions
    for $\qt>20~\GeV$.
    The horizontal error bars indicate the bin boundaries and the data
    points are positioned at the center-of-gravity of the bins based
    on the \POWHEG prediction.  The bands in the upper plot represent
    the uncertainty on the predictions from factorization and
    renormalization scales and PDFs.  The lower plot shows the ratio
    between the data and the theory predictions.  The bands in the
    lower plot represent the one standard deviation combined
    theoretical and experimental uncertainties.  }
    \label{fig:diffXsecResAccHighPt}
  \end{center}
\end{figure}

At high \qt, the precision of the prediction is dominated by the
perturbative order of the calculation and the handling of
the factorization and renormalization scale dependence.  In
Fig.~\ref{fig:diffXsecResAccHighPt} the measured normalized
differential distribution is compared to the prediction of \POWHEG  as
well as the ``Fully Exclusive \PW, \cPZ Production`` (\FEWZ)
package~\cite{FEWZ} for $\qt>20\GeV$ and $|\eta|<2.1$, calculated at
both O$(\alpha_s)$ and O$(\alpha_s^2)$.  The predictions were each normalized to their own
predicted total cross sections.  The \FEWZ calculation used the
effective dynamic scale definition $\sqrt{M_\cPZ^2+\left<\qt\right>^2}$
rather than the fixed scale of the \cPZ-boson mass. The
\FEWZ O$(\alpha_s^2)$ prediction produces a $\chi^2/\mathrm{ndof}$ of
30.5/9, which is a poorer agreement than the \POWHEG prediction (19.1/9),
particularly at the highest \qt.

\section{Summary}

Measurements of the normalized differential cross sections for
Drell--Yan muon and electron pairs in the \cPZ-boson mass region ($60
< M_{\ell\ell} < 120\GeV$) have been reported as functions of dilepton
rapidity and transverse momentum separately.  The results were
obtained using a data sample collected by the CMS experiment at the
LHC at a center-of-mass energy of 7\TeV corresponding to an integrated
luminosity of 36\pbinv.  The rapidity measurement is compared with the
predictions of several of the most recent PDF models and the agreement
evaluated as a function of the PDF set eigenvectors.  An overall
agreement between the models and the data is observed.  The measured
transverse momentum distribution is compared to various tunes of the
\PYTHIA generator for low transverse momentum and to O$(\alpha_s)$ and
O$(\alpha_s^2)$ predictions for high \qt.  No single model describes
the normalized differential cross section of the $\cPZ$ transverse
momentum over the full range.  These measurements significantly extend
previous Tevatron results and complement recent LHC results in
rapidity and transverse momentum.

\section*{Acknowledgments}

\hyphenation{Bundes-ministerium Forschungs-gemeinschaft
  Forschungs-zentren} We wish to congratulate our colleagues in the
CERN accelerator departments for the excellent performance of the LHC
machine. We thank the technical and administrative staff at CERN and
other CMS institutes. This work was supported by the Austrian Federal
Ministry of Science and Research; the Belgium Fonds de la Recherche
Scientifique, and Fonds voor Wetenschappelijk Onderzoek; the Brazilian
Funding Agencies (CNPq, CAPES, FAPERJ, and FAPESP); the Bulgarian
Ministry of Education and Science; CERN; the Chinese Academy of
Sciences, Ministry of Science and Technology, and National Natural
Science Foundation of China; the Colombian Funding Agency
(COLCIENCIAS); the Croatian Ministry of Science, Education and Sport;
the Research Promotion Foundation, Cyprus; the Estonian Academy of
Sciences and NICPB; the Academy of Finland, Finnish Ministry of
Education and Culture, and Helsinki Institute of Physics; the Institut
National de Physique Nucl\'eaire et de Physique des Particules~/~CNRS,
and Commissariat \`a l'\'Energie Atomique et aux \'Energies
Alternatives~/~CEA, France; the Bundesministerium f\"ur Bildung und
Forschung, Deutsche Forschungsgemeinschaft, and Helmholtz-Gemeinschaft
Deutscher Forschungszentren, Germany; the General Secretariat for
Research and Technology, Greece; the National Scientific Research
Foundation, and National Office for Research and Technology, Hungary;
the Department of Atomic Energy and the Department of Science and
Technology, India; the Institute for Studies in Theoretical Physics
and Mathematics, Iran; the Science Foundation, Ireland; the Istituto
Nazionale di Fisica Nucleare, Italy; the Korean Ministry of Education,
Science and Technology and the World Class University program of NRF,
Korea; the Lithuanian Academy of Sciences; the Mexican Funding
Agencies (CINVESTAV, CONACYT, SEP, and UASLP-FAI); the Ministry of
Science and Innovation, New Zealand; the Pakistan Atomic Energy
Commission; the State Commission for Scientific Research, Poland; the
Funda\c{c}\~ao para a Ci\^encia e a Tecnologia, Portugal; JINR
(Armenia, Belarus, Georgia, Ukraine, Uzbekistan); the Ministry of
Science and Technologies of the Russian Federation, the Russian
Ministry of Atomic Energy and the Russian Foundation for Basic
Research; the Ministry of Science and Technological Development of
Serbia; the Ministerio de Ciencia e Innovaci\'on, and Programa
Consolider-Ingenio 2010, Spain; the Swiss Funding Agencies (ETH Board,
ETH Zurich, PSI, SNF, UniZH, Canton Zurich, and SER); the National
Science Council, Taipei; the Scientific and Technical Research Council
of Turkey, and Turkish Atomic Energy Authority; the Science and
Technology Facilities Council, UK; the US Department of Energy, and
the US National Science Foundation.

Individuals have received support from the Marie-Curie programme and
the European Research Council (European Union); the Leventis
Foundation; the A. P. Sloan Foundation; the Alexander von Humboldt
Foundation; the Belgian Federal Science Policy Office; the Fonds pour
la Formation \`a la Recherche dans l'Industrie et dans l'Agriculture
(FRIA-Belgium); the Agentschap voor Innovatie door Wetenschap en
Technologie (IWT-Belgium); and the Council of Science and Industrial
Research, India.

\bibliography{auto_generated}   

\clearpage
\newpage

\cleardoublepage \appendix\section{The CMS Collaboration \label{app:collab}}\begin{sloppypar}\hyphenpenalty=5000\widowpenalty=500\clubpenalty=5000\textbf{Yerevan Physics Institute,  Yerevan,  Armenia}\\*[0pt]
S.~Chatrchyan, V.~Khachatryan, A.M.~Sirunyan, A.~Tumasyan
\vskip\cmsinstskip
\textbf{Institut f\"{u}r Hochenergiephysik der OeAW,  Wien,  Austria}\\*[0pt]
W.~Adam, T.~Bergauer, M.~Dragicevic, J.~Er\"{o}, C.~Fabjan, M.~Friedl, R.~Fr\"{u}hwirth, V.M.~Ghete, J.~Hammer\cmsAuthorMark{1}, M.~Hoch, N.~H\"{o}rmann, J.~Hrubec, M.~Jeitler, W.~Kiesenhofer, M.~Krammer, D.~Liko, I.~Mikulec, M.~Pernicka, B.~Rahbaran, H.~Rohringer, R.~Sch\"{o}fbeck, J.~Strauss, A.~Taurok, F.~Teischinger, C.~Trauner, P.~Wagner, W.~Waltenberger, G.~Walzel, E.~Widl, C.-E.~Wulz
% ...
\vskip\cmsinstskip
% ...
1:~~Also at CERN, European Organization for Nuclear Research, Geneva, Switzerland\\
% ...

\end{sloppypar}
\end{document}